\documentclass[prl,reprint, nofootinbib,amsmath,amssymb, aps, floatfix]{revtex4-1}
\usepackage{graphicx}%, xcolor, varwidth}
\usepackage{color}
\usepackage{braket}
\usepackage{mathrsfs}
\usepackage{amsthm}

\theoremstyle{remark}

\usepackage{bbold}
\usepackage{comment} 

\newcommand{\scri}{\mathscr{I}}
\begin{document}
\title{Firewalls From General Covariance}
\author{Raphael Bousso}
\affiliation{Center for Theoretical Physics and Department of Physics,\\
University of California, Berkeley, CA 94720, USA}

\begin{abstract}
%If the Almheiri \emph{et al.} (AMPS) thought experiment is performed, the horizon cannot be ``normal'': the predictions of semiclassical gravity coupled to effective field theory (SGEFT) must fail substantially. New physics beyond SGEFT must govern the horizon region. Proposals such as ER=EPR and nonisometric maps assert that this new physics will mimic horizon normalcy in a large class of global states, in that the black hole interior emerges nonlocally from the Hawking radiation.

%this new physics rules  ``horizon normalcy'' Here we discuss whether this new physics can mimic ``horizon normalcy,'' the predictions of 

I define ``horizon normalcy'' as the approximate validity of Semiclassical Gravity and Effective Field Theory (SGEFT), for the description of observers that approach or cross a black hole horizon. If black holes return information, then horizon normalcy must fail substantially, at least in some global states. Proposals such as ER=EPR and nonisometric maps assert that horizon normalcy persists, so long as the Hawking radiation remains in a computationally simple state. Here I argue that state-dependent horizon normalcy---independently of the underlying mechanism, and independently of the class of radiation states asserted to guarantee normalcy---requires a breakdown of general covariance far from the black hole, or else horizon normalcy will depend on the infinite future of the exterior. This is because the radiation can be in different states at different events, all spacelike to the horizon crossing event whose normalcy is at stake. I discuss a related effect in AdS/CFT, and I argue that its resolution by timefolds is of no help here.

%The simplest possibility, requiring nonlocal effects on the horizon scale $R$, is that it is impossible to enter a sufficiently old black hole. This is called a firewall. Alternatives such as ER=EPR or non-isometric maps appeal to nonlocality over the much greater scale $RS$ where the Hawking radiation is located; moreover they assert that the smoothness of the horizon is sensitive to the computational complexity of operations on the radiation. But at what \emph{time} should the state of the radiation be evaluated, to determine the experience of an observer crossing the horizon crossing at a specified event $E$? The radiation can be in both ``decoded'' and ``encoded'' states at different events, all spacelike to $E$ but eventually visible to the infalling observer. To single out one time would break general covariance in the distant radiation region; and to single out none would make the state at the horizon depend on the infinite future. I conclude that firewalls remain the least radical proposal for new physics.
\end{abstract}
\maketitle

\paragraph{\textbf{Firewall Problem}}

The AMPS firewall paradox~\cite{Almheiri:2012rt} can be formulated as follows~\cite{Almheiri:2013hfa}. A black hole is formed from a pure state and is allowed to evaporate fully. Assuming that ({\bf I}) this process is described by a unitary S-matrix, the entire Hawking radiation will be in some known pure state $\Psi$ after evaporation is complete. (Evidence for unitarity includes~\cite{Maldacena:1997re,Penington:2019npb,Almheiri:2019psf}. The alternative, that information is lost~\cite{Hawking:1976ra}, will not be explored here.)

\begin{figure}
\includegraphics[width=0.47\textwidth]{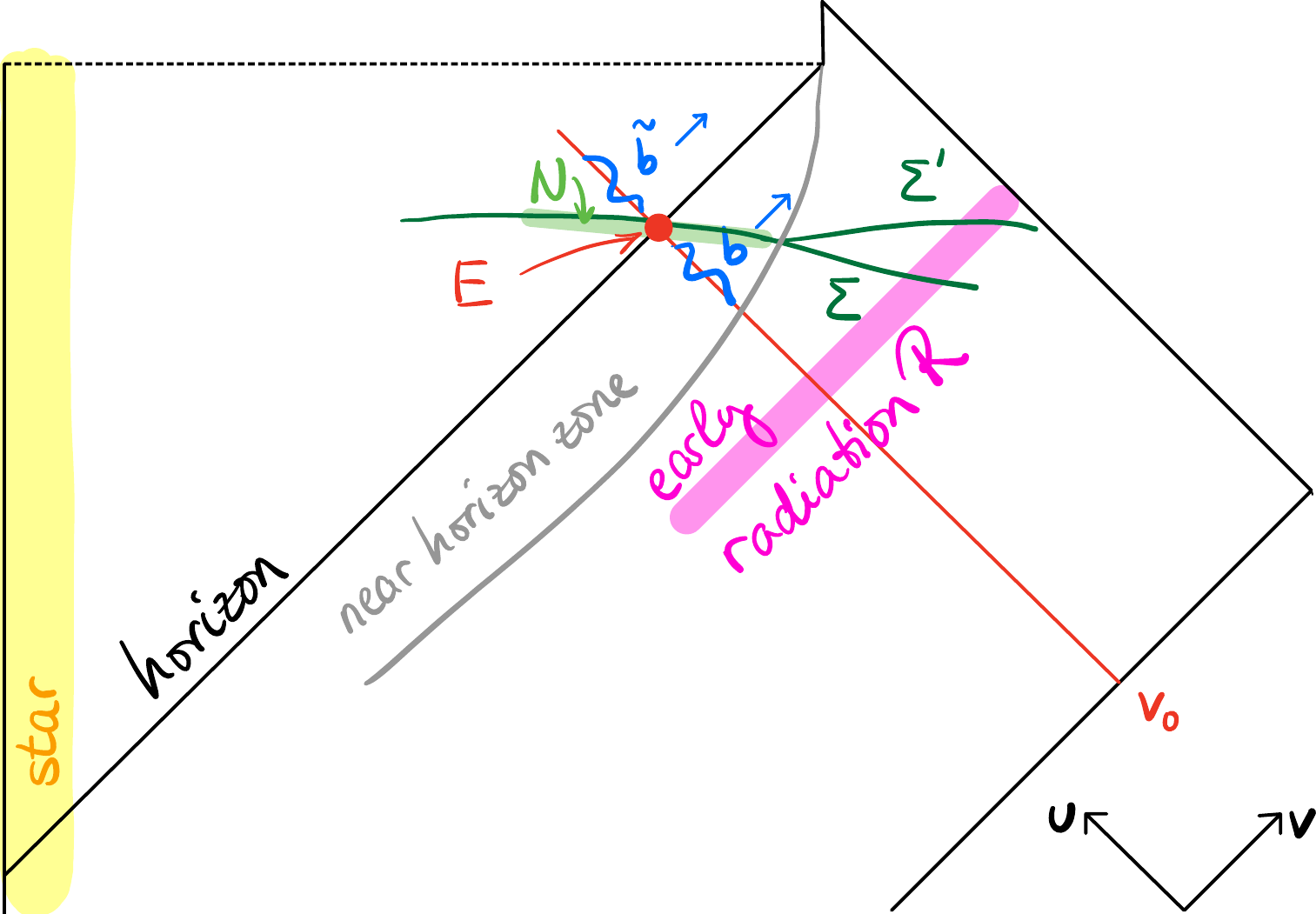}
\caption{\label{fig-horizon}
{\footnotesize Penrose diagram of an evaporating black hole. $E$ is a fixed event at which Alice plans to cross the horizon. Here we examine the hypothesis~\cite{Maldacena:2013xja,Akers:2022qdl} that horizon normalcy (the unambiguous validity of SGEFT in a local neighborhood $N$ of $E$ probed by Alice) depends on the state of distant Hawking radiation that has left the near horizon zone before $v=v_0$. Since the radiation state can differ at different times $\Sigma$, $\Sigma'$, this hypothesis requires either the introduction of a preferred global time slice, or else the dependence of horizon normalcy at $E$ on the infinite future of the exterior.}}
\end{figure}

Consider an outgoing Hawking particle $b$ in a neighborhood of a cut $v=v_0$ of the black hole horizon; see Fig.~\ref{fig-horizon}. For a Schwarzschild black hole of radius $r_S$, the near-horizon zone is the region $0<r-r_S\ll r_S$. In this zone, the Hawking radiation modes are wave-packets constructed from Rindler-like modes, with the black hole horizon playing the role of the Rindler horizon~\cite{Unruh:1976db}. For definiteness, let $b$ be a spherically symmetric wavepacket, with radial support more than a Planck distance away from the horizon but with radial width much less than the black hole radius. We work in Eddington-Finkelstein coordinates,
\begin{equation}
    ds^2 = -\left(1-\frac{r_S}{r}\right) dv^2 - 2\,dv\,dr + r^2 d\Omega^2~.
\end{equation}

Now assume that ({\bf II}) the evolution of $b$ from the near-horizon region to the asymptotic region is governed by effective field theory. By assumptions ({\bf I,II}), the outgoing mode $b$ will be entangled with and purified by the rest of the Hawking radiation in the state $\Psi$. (This is true in a typical state $\Psi$. If $b$ is pure or less than typically entangled, the problem is worse~\cite{Bousso:2013wia}.) 

Not all of the Hawking radiation in $\Psi$ has been emitted yet at the time $v=v_0$. However, at any time $v_D>v_P$, it is possible~\cite{Bennett:1999hf,Hayden:2007cs} to distill a nearly complete purification $e_b$ of $b$ from the ``early'' radiation $R$ that has already come out by that time. Here $v_P\sim S_{\rm BH} r_S$ is the Page time, when the coarse-grained Hawking radiation entropy begins to exceed the Bekenstein-Hawking entropy of the black hole. That is, there exists a (non-unique) unitary $U$ acting on $R$ at the time $v_D$ that outputs a particle $e_b$ such that $be_b$ are together in a nearly pure state, tensored with some state for the remaining radiation system $R\setminus e_b$. The deviation from purity, the entropy $S_{be_b}$, will be exponentially small in the number of Hawking particles emitted between $v_P$ and $v_D$. We will take this number to be large, $T_H(v_D-v_P)\gg 1$, where $T_H\sim r_S^{-1}$ is the Hawking temperature. 

%We may choose $v_0>v_D$, so that the distillation of $e_b$ precedes the time when the Hawking particle $b$ is near the horizon. Thus, 
If $v_D<v_0$,
an observer Alice can put $e_b$ in her pocket and schedule her jump so as to encounter $b$ at $v=v_0$ near the horizon. By assumptions ({\bf I,II}), Alice must find that $b$ and $e_b$ together are in the pure state predicted by unitarity. (Alice can verify this before she crosses the horizon, while she could still decide to stay outside and wait for the black hole to evaporate fully. This is important because black hole complementarity~\cite{Susskind:1993if} allows for apparent contradictions between an infalling observer's findings and the exterior description of a black hole, so long as the conflict cannot be experienced by or communicated to any single observer. Under this paradigm, if Alice were able to jointly probe $b$ and $e_b$ only \emph{after} entering the black hole, she need not find them in the state demanded by unitarity.)

Let us now assume ({\bf III}) \emph{horizon normalcy}. Let $N$ be a neighborhood of the event $E$ of characteristic size $r_S$ that excludes the high-curvature region near the singularity, but includes the near-horizon zone. Horizon normalcy is the statement that $N$ is described by SGEFT, up to nonperturbative corrections. 

Horizon normalcy implies that the horizon itself looks locally like the vacuum~\cite{OppenheimerSnyder}. In the vacuum state, $b$ is strongly entangled with and purified by a ``partner mode'' $\tilde b$, which is also outgoing and localized just inside the horizon at the time $v_0$~\cite{Hawking:1975vcx}. (This step uses the fact that $b$ is localized in the near-horizon region.)

To summarize: by ({\bf III}), horizon normalcy, the von Neumann entropies satisfy $S_{\tilde b b}\ll 1$ and $S_b\sim O(1)$; by ({\bf I,II}), unitarity, $S_{be_b}\ll 1$. But strong subadditivity~\cite{Lieb:1973cp} implies $S_{\tilde b b}+S_{be_b}\geq S_b$~\cite{Mathur:2009hf}. Thus Alice has local access to a tripartite system in a mathematically impossible quantum state. It follows that assumptions ({\bf I,II,III}) cannot all be true.\footnote{This conclusion is independent of the underlying quantum gravity theory~\cite{Mathur:2012jk} from which SGEFT might emerge; it implies that ({\bf I,II,III}) cannot all emerge. I omit a host of potential subtleties and suspected loopholes that were explored in the early literature on the firewall paradox (see Ref.~\cite{Almheiri:2013hfa} for a discussion and references). These include the hypothesis that the decoding unitary cannot be implemented fast enough, as a matter of physical law rather than mere practicality~\cite{Harlow:2013tf}. To my knowledge, this idea was not developed further; perhaps it should be.} 

If we commit to assumptions ({\bf I}) and ({\bf II}), then it follows that Alice will not see a normal horizon ({\bf III}).  This argument applies to any outgoing Hawking mode near the horizon, including higher angular momentum modes, limited only by some high-energy cutoff. Thus the infalling observer should encounter a violent ``firewall''~\cite{Almheiri:2012rt}. 

Our focus will be on horizon normalcy, not on the violence of what might replace it. From a principled standpoint, any easily detectable departure from normalcy is as radical as a firewall. This includes the presence of just a single near-horizon mode $\tilde b b$ whose occupation number differs from the Unruh state, or the ``nonviolent nonlocality'' models~\cite{Giddings:2012gc,Giddings:2024qcf}. I will examine whether horizon normalcy can be recovered for a class of global states.%, including situations where the AMPS experiment is not performed.

\paragraph{\textbf{State-Dependent Horizon Normalcy}}

The AMPS thought experiment is compelling because it is operational. When Alice crosses the horizon, she has local, simultaneous access to all three relevant particles $\tilde b, b, e_b$. Assuming ({\bf I,II}), the fact that the horizon is abnormal (i.e., that at least $\tilde b b$ differs substantially from the vacuum state) is then beyond dispute. 

In standard quantum mechanics, this conclusion holds regardless of whether anyone actually extracts $e_b$ before Alice jumps in. Even if $e_b$ remains scrambled in the radiation, the fact that it nearly purifies $b$ makes it inconsistent for $b$ to be purified by $\tilde b$ in the vacuum state. 

Unless the purification $e_b$ is decoded into a small system, however, this contradiction will not manifest itself directly to any one observer. (Alice cannot transport the entire early Hawking radiation to the horizon without creating a much larger black hole.) Moreover, the extraction of $e_b$ is computationally hard: it requires a number of simple logical gates that scales exponentially with the area of the remaining black hole~\cite{Harlow:2013tf}. 

Based on these observations, it has been proposed that \emph{the horizon will remain normal if the state of the Hawking radiation remains in a certain class of states}---for example, so long as no complex operation is performed on the Hawking radiation. This proposal implies nonlocality on the distance scale $r_S S_{\rm BH}$ accessed by the Hawking radiation, which is much greater than that of the black hole. In addition, it requires---and so far, lacks---a precise definition of the normalcy-guaranteeing class of states. (E.g., what counts as a ``simple'' operation? The rigorous definition of complexity classes takes the input size of the problem to infinity. Real-world settings are incompatible with this scaling, so a complexity class would have to be assigned at finite input size.) Moreover, one would need to keep track of the complexity of the ``Hawking radiation'' even if the latter is absorbed by other systems that could already be arbitrarily complex themselves.

We ignore these challenges here, in order to exhibit a robust, universal problem: state-dependent horizon normalcy requires a preferred global time slice, and thus a violation of general covariance far from the black hole. Without a preferred time, horizon normalcy will depend on the infinite future---a form of nonlocality arguably to extreme to contemplate.

\paragraph{\textbf{Time Ambiguity}}

We begin by formalizing the idea of state-dependent horizon normalcy. 
%We will then show that it is inconsistent with general covariance in weakly gravitating regions far from the black hole. Later, we will turn to more specific proposals such as ER=EPR, but the argument against the dependence of horizon normalcy on the global state is quite general.
Let $\Sigma$ be a global Cauchy slice that contains $E$, and let $\rho_N$ be the reduced state of the quantum fields in the open neighborhood $N$ of $E$ on $\Sigma$ in which horizon normalcy should hold, extending less than $S_{\rm BH} r_S$ from the horizon. Let $\rho_\text{rad}^\text{enc}$ (``encoded'') be the state of the radiation on $\Sigma$ reached by the formation and natural evaporation of a black hole, at a distance of order $r_S S$ from the horizon. Let $\rho_\text{rad}^\text{dec}$ (``decoded'') be a radiation state on $\Sigma$ in which $e_b$ has been decoded. To obtain the state $\rho_\text{rad}^\text{dec}$ on $\Sigma$, some agent at an earlier time will have distilled the purification of $b$ from the early radiation, so that $e_b$ is available on $\Sigma$ as a physical localized particle. 

If the radiation state on $\Sigma$ is $\rho_\text{rad}^\text{dec}$, then the AMPS argument precludes horizon normalcy: $\rho_N$ must contain an excitation (at least) in the infalling mode built from $b$ and $\tilde b$. But if the radiation state on $\Sigma$ is $\rho_\text{rad}^\text{enc}$, then the above proposal asserts normalcy, i.e., it asserts that $\rho_N$ is the vacuum state reduced to a neighborhood of $E$.

Now let $\Sigma'$ be another Cauchy slice that coincides with $\Sigma$ on $N$ (and which therefore also contains $E$), but which meets the Hawking radiation at a different time; see Fig.~\ref{fig-horizon}. Thus the (Schr\"odinger) state on $\Sigma'$ can be different from that on $\Sigma$. For example, the radiation on $\Sigma$ might be in the state $\rho_\text{rad}^\text{enc}$; but later $e_b$ is decoded, so that on $\Sigma'$ the state is $\rho_\text{rad}^\text{dec}$. Or conversely, $e_b$ could have been decoded on $\Sigma$ but then re-encoded into the radiation by acting with $U^\dagger$, so that $\Sigma'$ contains the state $\rho_\text{rad}^\text{enc}$.

If horizon normalcy near $E$ depends on the state of the radiation, then it is not clear whether it should be determined from the radiation state on $\Sigma$, or on $\Sigma'$, or perhaps on yet another time slice. The only covariant choice is the boundary of the past of $E$, $v=v_0$ in our example. But this choice fails~\cite{Almheiri:2013hfa}: suppose the state at $v=v_0$ is $\rho_\text{rad}^\text{enc}$, leading to horizon normalcy at $E$. Then if $e_b$ is decoded within a scrambling time $T_H^{-1}\log S_{\rm BH}$ after $v_0$, it can be sent into the black hole and received by Alice, leading again to an impossible quantum state of the tripartite system $\tilde b b e_b$ locally accessible to Alice.

Therefore, horizon normalcy cannot depend on the state of the radiation, unless a preferred time slice is introduced far from the black hole on which that state is to be considered. But this would violate general covariance far from the black hole, in a distant and weakly gravitating region. This seems far more radical than a breakdown of SGEFT only near the black hole, on a distance scale comparable to the black hole radius.

The remaining alternative is that horizon normalcy at $E$ is guaranteed only if the radiation state remains in the horizon-normalcy-guaranteeing class (here, ``encoded'') \emph{forever}.\footnote{Some approaches (including those discussed below) define $e_b$ as ``encoded'' so long as the computational complexity of decoding $e_b$ remains in the exponential class identified by Harlow and Hayden~\cite{Harlow:2013tf}. By this definition, $e_b$ must be regarded as ``automatically decoded'' after the black hole has fully evaporated: at this time, the task may be arduous but it is no longer exponentially hard. With this definition, a nontrivial dependence of horizon normalcy on the infinite future is not an option, because the radiation state always exits the normalcy-guaranteeing class at sufficiently late times. The choice is between introducing a preferred time in the radiation region and thus violating general covariance, or giving up on state-dependent horizon normalcy.} A dependence of normalcy at $E$ on the infinite future of the exterior should again be contrasted with the comparatively mild nonlocality of a firewall.

\paragraph{\textbf{ER=EPR and Non-Isometric Maps}}

The time ambiguity problem exhibited above arises in any proposal in which horizon normalcy depends on the state of the distant Hawking radiation. It is irrelevant \emph{why} horizon normalcy depends on the radiation state. It also does not matter whether the class of states sufficient for normalcy is characterized by the absence of complexity, or by some other criterion. Nevertheless, it is instructive to pinpoint how time-ambiguity shows up in approaches that invoke a specific mechanism. 

I will focus on two ideas that have received significant attention: ER=EPR~\cite{Maldacena:2013xja},\footnote{See Refs.~\cite{Bousso:2012as,Verlinde:2013qya} for related ideas, and Refs.~\cite{Bousso:2012as,Almheiri:2013hfa,Bousso:2013wia,Bousso:2013ifa,Marolf:2013dba,Harlow:2014yoa,Marolf:2015dia}, for criticisms unrelated to the time ambiguity emphasized here.} and non-isometric maps~\cite{Akers:2022qdl}. Both assert horizon normalcy if the radiation state lies in some restricted class, and thus both suffer from the time ambiguity.
This is obscured, however, when we rely on toy models that capture key elements of each idea, but which omit too much of the physical spacetime structure of a real black hole to capture the time ambiguity.

ER=EPR invokes an analogy with a two-sided black hole in the Hartle-Hawking (or thermofield-double) state, which is assumed to have a smooth horizon. (This assumption is not trivial~\cite{Marolf:2012xe}.) Assuming a typical entanglement pattern between the physical black hole and the radiation it has emitted by the time $v_0$, there exists a unitary acting on the radiation that will convert the radiation into a second black hole of equal size, plus leftover radiation, with the two black holes in the thermofield-double state. Essentially, ER=EPR is the conjecture that this operation happens automatically as far as the infalling observer Alice is concerned, guaranteeing horizon normalcy. An exception has to be made if $e_b$ has been decoded from the radiation, because then the horizon cannot be normal. This case is claimed~\cite{Maldacena:2013xja} to be analogous to modifying the Hartle-Hawking state by removing or altering the mode $\tilde b$ when it is a simple physical wavepacket near the left (unphysical) boundary of the two-sided black hole. 

But the decoding of $e_b$ from the early Hawking radiation $R$ is never simple. As a complex operation, it can be done, and undone, at various events that are all spacelike related to the horizon crossing event $E$. No criterion is available analogous to assessing whether $\tilde b$ is in the correct state on the left boundary at the (preferred) time when its state could be altered by a simple operator.

Let us turn to non-isometric maps~\cite{Akers:2022qdl}, which implement state-dependent horizon normalcy as follows. One considers the fundamental state $\Psi$ of the radiation and black hole (analogous to the boundary state in AdS/CFT or the final state of the radiation arriving at $\scri^+$ in asymptotically flat spacetime). One asks if $\Psi$ is the image of a ``simple'' effective global state $\psi$ of the semiclassical geometry, under a non-isometric map $V$. If the answer is yes, then $\rho_N$ is determined by reducing $\psi$ to $N$. This ensures horizon normalcy if the Hawking radiation was not subjected to complex operations. If the answer is no, then horizon normalcy may fail in $N$. In particular, the answer will be no if $e_b$ has been decoded from the radiation: since decoding requires a complex unitary, $\Psi$ is then not the image of a ``simple'' effective state.

As in ER=EPR, the global state that is determinative for normalcy is implicitly assumed to be unique and definite: the state is either ``simple,'' or not. But for a real black hole, we have seen that both statements can be true, on different Cauchy slices that all include the same neighborhood $N$ of $E$. The key limitation of the toy model in Ref.~\cite{Akers:2022qdl} is that it fails to capture this aspect of causal structure (rather than, for example, that the map $V$ was constructed only for a qudit model).

\paragraph{\textbf{Entanglement Wedge Conflict and Timefolds}}

\begin{figure}
    \centering
    \includegraphics[width=0.45\textwidth]{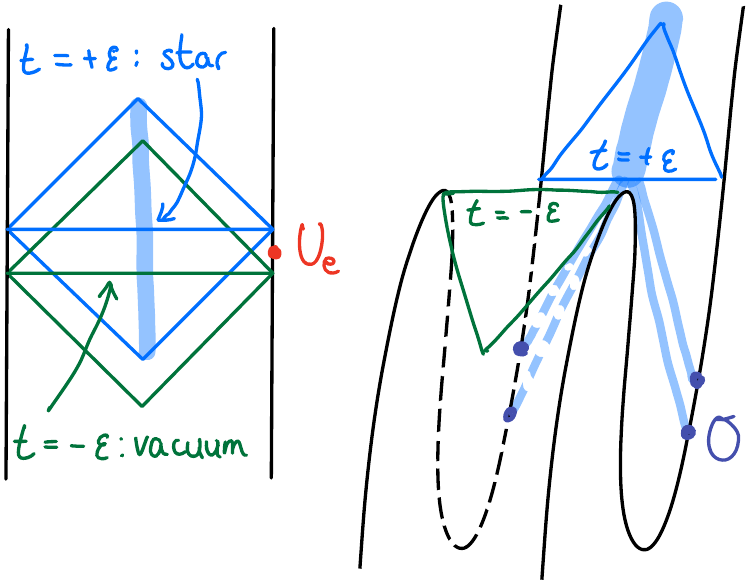}
    \caption{Left: entanglement wedge conflict in AdS/CFT. At the boundary time $t=0$, the boundary operator $U_e$ changes the CFT vacuum to a state dual to a star. The Wheeler-de Witt patches of the boundary times $t=\pm \epsilon$  (diamonds) overlap inconsistently in the bulk. Right: by inserting a timefold, the conflict is mitigated. The star can be built by sending in excitations from the boundary using local operators $\cal O$. There are two copies of the time $t=-\epsilon$, one in the vacuum state (green horizontal line), and one with a nearly complete star.}%---In black holes formed by collapse, timefolds are not predicted by SGEFT and hence conflict with horizon normalcy.}
    \label{fig:timefolds}
\end{figure}

The time ambiguity problem can be viewed as a special case of \emph{entanglement wedge conflict}, a broader limitation on bulk emergence~\cite{Bousso:2012sj,Bousso:2023sya}; see Fig.~\ref{fig:timefolds}. Consider a holographic CFT on in its vacuum state. At $t=0$, act with a unitary $U_e$ that maps the vacuum to an excited state, containing for example a star in the center of the bulk. $U_e$ is nonlocal as a CFT operator. ($U_e$ is not complex in the sense of Ref.~\cite{Harlow:2013tf}, however, since $U_e$ arises by Hamiltonian evolution of local CFT operators over a polynomial time~\cite{Hamilton:2006az}.) The entanglement wedge of any constant boundary time slice $t\neq 0$ is a well-defined Wheeler-de Witt patch containing the entire bulk, either in the vacuum ($t<0$), or with a particle ($t>0$). But the entanglement wedges of $t=\pm \epsilon$ overlap for small $\epsilon$ and so are mutually inconsistent. This is inevitable, since there is no preferred bulk time spacelike to the $t=0$ boundary slice when the particle should appear. Indeed, the sudden appearance of an excitation deep in the bulk, spacelike to $U_e$, would violate the Bianchi identity.

It is possible to mitigate entanglement wedge conflict by introducing ``timefolds''~\cite{Heemskerk:2012mn}. The idea is to cut the boundary at $t=0$ and ``add in'' enough time for at least 2 light-crossings of the bulk. Time runs backwards on the first half of the new piece, both on the boundary and in the bulk. Time runs forward in the second half. This makes it possible to build up the new state at $t=\epsilon$ using local boundary operators $\cal O$ only. The length of the timefolds and the operators $\cal O$ are highly non-unique.

Now let us return to the example of an evaporating black hole. The ``time ambiguity'' we exhibited has some commonalities with entanglement wedge conflict in AdS/CFT. The early Hawking radiation $R$ plays the role of the CFT. The two states at $t=\pm\epsilon$ are analogous to our earlier $\rho_{\rm dec}$ and $\rho_{\rm enc}$. After the Page time, the entanglement wedge of $R$ contains an ``island'' $I(R)$ that consists of the black hole interior a scrambling time  earlier~\cite{Penington:2019npb,Almheiri:2019psf}. The ability to decode $e_b$ from the radiation is related in a precise way to the fact that $I(R)$ contains the mode $\tilde b$; this ensures that Petz reconstruction of operators acting on $\tilde b$ succeeds~\cite{Penington:2019kki,Almheiri:2019qdq}. 

Given these parallels, let us ask whether timefolds can mitigate the time ambiguity problem, by preserving general covariance in the radiation region and instead introducing multiple copies of the event $E$ on different timefolds in the black hole region: one for each decoding or re-encoding operation ever performed on the radiation. 

This would require a generalization of timefolds to entanglement islands. At present there is no such proposal, perhaps because the analogy with the AdS/CFT setting is imperfect: 1.~The proposals~\cite{Maldacena:2013xja,Akers:2022qdl} assert that acting on \emph{all} of the early radiation $R$ with a \emph{simple} operator will have no effect at the horizon, but certain complex operators will destroy horizon normalcy. By contrast, the AdS-timefold trigger $U_e$ is simple. Moreover, almost every operator acting on \emph{all} of the same bulk degrees of freedom, simple or complex, will have large effects in the bulk. 2.~The conformal boundary of AdS anchors global bulk time-slices that play the role of the turning times of the bulk timefolds. By contrast, the decoding and encoding events in the Hawking radiation take place far from the island. State-dependent horizon normalcy requires these events to be interpreted as inserting timefolds in the black hole region but not in the far region. 
%No-one controls or manipulates the sign of the Hamiltonian in this weakly gravitating region, so the timefolds should not extend to it. 
It is not clear what bulk region near or at the black hole should be subjected to timefolding, or how to connect these timefolds to the rest of the spacetime. %This issue is related to another important difference compared to inserting $U_e$ in the CFT. 
%3.~It is not clear whether $\rho_{\rm dec}$ admits any semiclassical description of the black hole interior.

\begin{figure}
\includegraphics[width=0.47\textwidth]{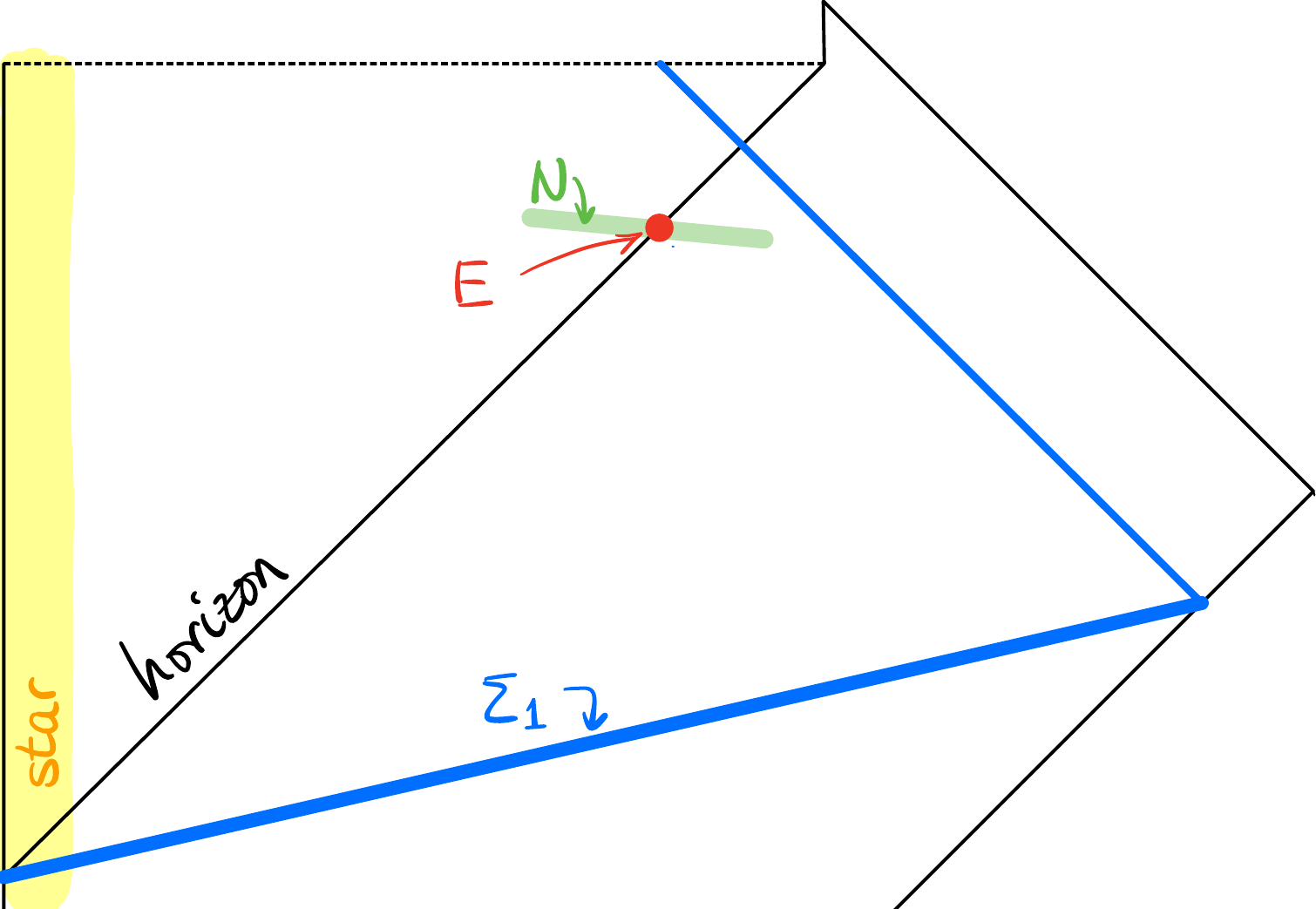}
\caption{\label{fig-notimefold}
{\footnotesize Horizon normalcy requires the validity of SGEFT. Since $N$ lies inside the domain of dependence of $\Sigma_1$ (blue wedge) according to SGEFT, the data on the partial Cauchy slice $\Sigma_1$ must evolve into the unique state in the region $N$ predicted by SGEFT. This is inconsistent with timefolds, which assign different states to multiple copies of $N$. By contrast, the 
%timefold in Figure 2 is associated with evolution beyond the Cauchy horizon of SGEFT. (AdS boundary conditions that permit the operator $U_e$ preclude bulk global hyperbolicity.) The
evolution from $t=-\epsilon$ to $t=+\epsilon$ in Fig.~2 is under control through the CFT and transcends SGEFT in the bulk. It can be given a bulk interpretation as a timefold; but there is no analog of the requirement of horizon normalcy, and hence no need for SGEFT to determine a unique bulk evolution between $t=\pm \epsilon$.}}
\end{figure}
We can put these unsolved issues aside, because timefolds would not restore horizon normalcy in any case. The question we set out to answer was: can horizon normalcy depend on the state of the radiation? Horizon normalcy requires the unambiguous validity of the standard predictions of SGEFT for the experiences of an observer crossing the horizon. \emph{Timefolds are not predicted}~\cite{OppenheimerSnyder}, so horizon normalcy requires their absence; see Fig.~\ref{fig-notimefold}. This in turn requires that no decoding operation will \emph{ever} be performed on the radiation, again leading to a dependence of horizon normalcy on the infinite future.

\paragraph{\textbf{Conclusion}} 

For black holes to return information, new physics is needed at the horizon of a sufficiently old black hole. State-dependent horizon normalcy asserts that the new physics conspires to avoid detectability in some class of global states. I have shown that this necessitates a preferred time slicing far from the black hole; or else, horizon normalcy depends on the infinite future.

%This is true even if horizon normalcy were possible in some class of states. And the new physics (as a theory, if not as a phenomenon) is needed \emph{even in these hypothetical, preferred states,} because the smooth black hole horizon, unlike other surfaces, would emerge nonlocally from the distant Hawking radiation. 
%Models such as ER=EPR accept that new physics is required at the horizon, but then work hard to recover the old physics at the horizon: its normalcy. 

%But why would new physics cover its tracks? 

We have evidence for general covariance in weakly gravitating regions, but not for horizon normalcy, so %Here I have argued that state-dependent horizon normalcy is incompatible with general covariance far from the black hole. 
it seems more likely that the new physics does not hide itself. Firewalls involve nonlocality ``only'' on the scale of the black hole. By conceding a universal failure of horizon normalcy, the firewall paradigm avoids the breakdown of SGEFT in weakly gravitating regions, or even infinite nonlocality. 
%They allow SGEFT to remain a good approximation far from the black hole. 

However, there is currently no proposal for an effective description of firewalls and their dynamics. They may be as far as $(Gr_S^2)^{1/4}$ from the horizon~\cite{Bousso:2023kdj} in stationary black holes. A covariant criterion for their location will be suggested in forthcoming work~\cite{Bousso:2025toappear}.

\paragraph{\textbf{Acknowledgements}}
I am indebted to C.~Akers, A.~Almheiri, N.~Engelhardt, D.~Harlow, J.~Maldacena, G.~Penington, A.~Shahbazi-Moghaddam, D.~Stanford, L.~Susskind, and M.~Usatyuk for extensive discussions. This work was supported by the Department of Energy under QuantISED Award DE-SC0019380.

\bibliographystyle{JHEP}
\bibliography{bibliography}

%\bibliographystyle{utcaps}
%\bibliography{main}

\end{document}